\documentclass[final,3p,times,twocolumn]{elsarticle}

\usepackage{graphicx}

\usepackage{amssymb}
\usepackage[english]{babel}

\begin{document}

\begin{frontmatter}



\title{The MRPC--based ALICE Time--Of--Flight detector: status and performance}


\author {A. ALICI (for the ALICE Collaboration)}

\address {Centro Studi e Ricerche e Museo Storico della Fisica "Enrico Fermi", Rome, Italy\\
Dipartimento di Fisica dell'Universit\`a and Sezione INFN, Bologna, Italy\\
E-mail: \texttt{alici@bo.infn.it}}

\begin{abstract}
The large Time--Of--Flight (TOF) array is one of the main detectors devoted to charged hadron identification in the mid--rapidity region of the ALICE experiment at the LHC. It allows separation among pions, kaons and protons up to a few GeV/c, covering the full azimuthal angle and -0.9 $< \eta <$ 0.9. The TOF exploits the innovative MRPC technology capable of an intrinsic time resolution better than 50 ps with an efficiency close to 100\% and a large operational plateau; the full array consists of 1593 MRPCs covering a cylindrical surface of 141 m$^2$.\\
The TOF detector has been efficiently taking data since the first pp collisions recorded in ALICE in December 2009. In this report, the status of the TOF detector and the performance achieved for both pp and Pb--Pb collisions are described.

\end{abstract}



\end{frontmatter}


\section{Introduction}
\label{intro}




ALICE \cite{alice} (A Large Ion Collider Experiment) is the dedicated heavy--ion experiment at the CERN LHC focusing on the study of the physics of strongly interacting matter at extreme energy densities, where the formation of a new phase of matter, the quark--gluon plasma (QGP), is expected. ALICE is also studying proton--proton collisions both for comparison with Pb--Pb collisions and in physics areas where ALICE is competitive with other LHC experiments. Particle IDentification (PID) is a crucial aspect of the ALICE experiment. The silicon Inner Tracking System (ITS) and the Time Projection Chamber (TPC) identify charged particles below 1 GeV/c using the $dE/dx$ measurements. From 1 GeV/c to a few GeV/c the Time--Of--Flight (TOF) detector extends the PID capabilities of ALICE with a complete coverage of the central region. A Transition Radiator Detector (TRD) and a Cherenkov Ring Image detector (HMPID) further increase the PID capabilities of the experiment. A detailed review of ALICE and of its PID strategy can be found in \cite{pid}.\\
\indent The TOF \cite{tof} is a large area detector covering a cylindrical surface of 141 $\mbox{m}^2$ with an inner radius of 3.7 m, a pseudorapidity interval [-0.9,+0.9] and full azimuthal coverage. The TOF exploits the innovative Multigap Resistive Plate Chamber (MRPC) \cite{mrpc} technology, capable of an intrinsic time resolution better than 50 ps with an efficiency close to 100\% and a large operational plateau. The whole system is made of 1593 MRPCs arranged into 90 gas--tight modules which are grouped into 18 SuperModules (SM) each covering an azimuthal angle of 20 degrees.

\section{The ALICE TOF MRPCs}
\label{TOFmrpc}

Unlike the standard RPCs, in the MRPCs the gas gap between the electrodes is divided by means of internal plates which are physical barriers stopping the avalanche growing too big; in this way it is possible to apply a very intense electric field and still operate the device in avalanche mode with many advantages in terms of time resolution and rate capability. The very good time resolution of the MRPC is due to the strong uniform electric field, which provokes the avalanche process immediately after primary ionization is deposited in the gas volume. As a consequence, the intrinsic detector resolution is determined mainly by the avalanche statistics.\\
\indent The MRPCs for the ALICE TOF are designed as double--stack strips. Each stack has 5 gas gaps of 250 $\mu$m width; the two stacks are placed on each side of a central anode. The resistive plates are made of high--resistivity ($\approx$$10^{13}$ $\Omega$cm) \emph{soda--lime} commercial glass 0.4 mm thick. Compared to melamine--phenolic plastic materials, the choice of glass plates ensures higher uniformity of the gap size and higher stability for long--term behavior. On the outer surface of the external plates a very intense electric field ($\simeq$100 kV/cm) is applied by means of a specially developed acrylic paint loaded with metal oxides giving a surface resistivity between 2 and 25 M$\Omega/\square$. The pick--up elements are 2.5 $\times$ 3.5 $\mbox{cm}^2$ area pads placed outside the external plates. The gas mixture used is C$_2$H$_2$F$_4$ (93\%) and SF$_6$ (7\%).\\
\indent Test beam results may be found in \cite{test_beam,constr}, where a time resolution better than 50 ps, including all of the electronic readout contribution, and an efficiency close to 100\% with a long streamer--free efficiency plateau have been measured.


\section{HV, noise and trigger}
\label{commi}

The commissioning of the full TOF detector was performed in 2008 and 2009 \cite{2008,2009} and all operating parameters have been measured to be well within the expectations of the project.\\
\indent Measurements with the HV system indicate that without beam the dark current of the ALICE TOF MRPCs is extremely low, at the nA/MRPC level at 13 kV (6.5 kV positive and 6.5 kV negative HV). During LHC operation, even at the maximum charged particle flux sustainable by the ALICE detector, corresponding to a istantaneus luminosity of $L$ = 2$\cdot$10$^{30}$s$^{-1}$cm$^{-2}$, the current drawn by the 1593 MRPCs comprising the TOF is around 35 $\mu$A, less than 25 nA per MRPC. This very small current indicates that no aging effect are present in the MRPCs.\\
\indent  Also detector noise benefits from this small current. The TOF dark noise has been found substantially reduced with respect to predictions; at the nominal threshold an average rate of 0.2 Hz/cm$^2$ has been measured. Thus, the TOF trigger is the preferred choice for cosmic runs in ALICE, with a rate of $\approx$ 10 Hz for pure back--to--back topology and 50\% of purity \cite{2009}. We note that these results have been achieved running all the MRPCs at the same voltage and with all front--end electronics thresholds set at the same value.\\
\indent During the LHC heavy--ion program in November 2010 the TOF contributed to the trigger of Ultra--Peripheral Collisions (UPC) in the barrel. In such events two relativistic nuclei collide with impact parameter larger than twice their radius; they provide a powerful tool to study the gluon distribution function in the nuclei without hadronic background. A relevant process in UPC is the production of vector mesons containing heavy flavours. The total cross section for this process is a small fraction of the total Pb--Pb cross section, therefore dedicated triggers have been used to tag these events both in the barrel and in the forward muon arm of ALICE. The UPC trigger in the barrel requires signals coming from the central detectors (ITS and TOF) with forward detectors not showing any activity. In Fig. 1 $J/\psi$ candidates from coherent photoproduction (characterised by low transverse momentum of the final states $p_t <$ 200 MeV/c) in Ultra--Peripheral Pb--Pb collisions at $\sqrt{s_{NN}}$ = 2.76 TeV are shown. With the full 2010 statistics more than 10000 $\rho^0$ and $\approx$ 40 $J/\psi$ candidates have been found.

\begin{figure}
\label{upc}
\includegraphics[width=7.5cm, height=7.5cm]{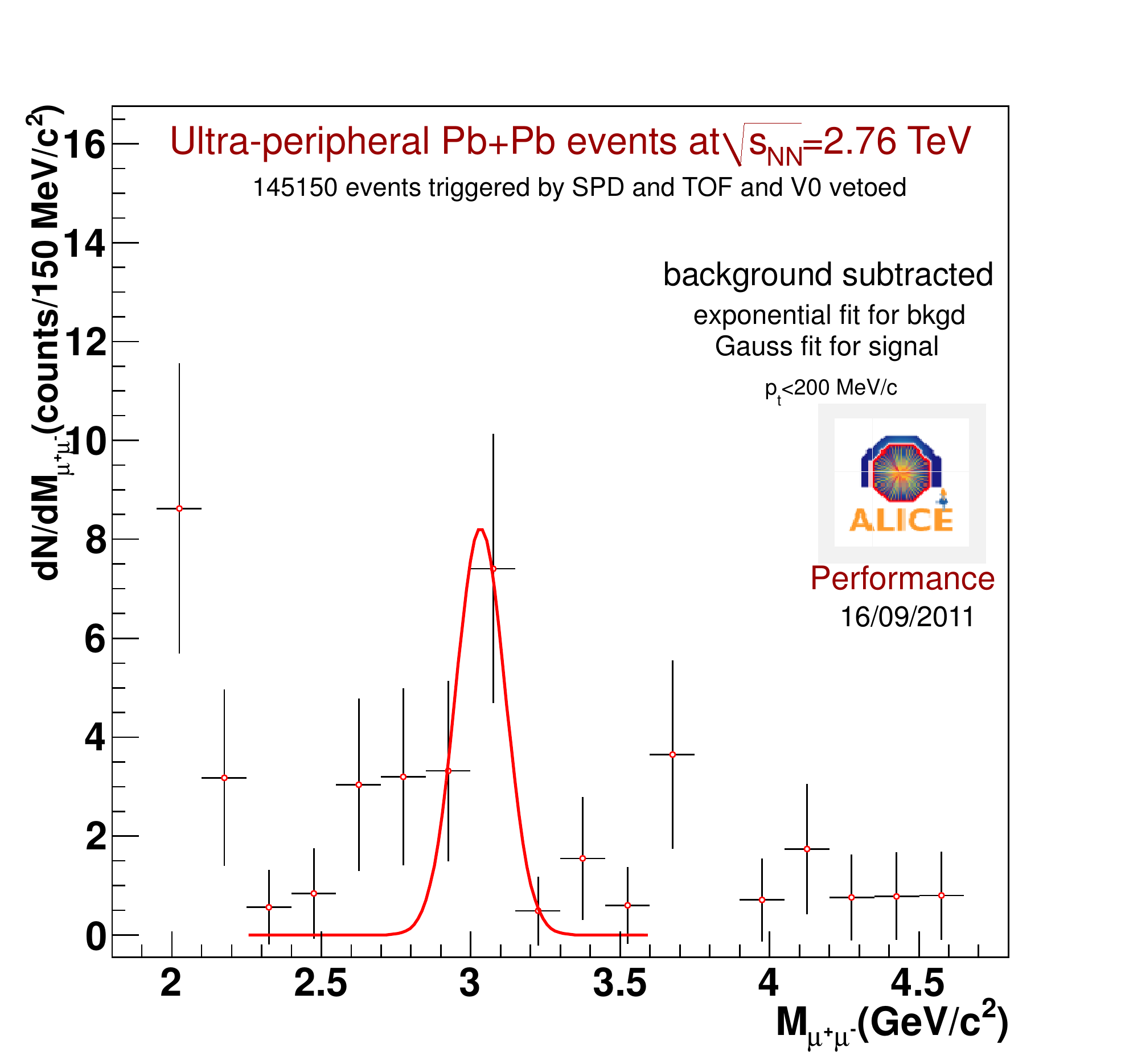}
\caption{Invariant mass for $J/\psi$ in UPC assuming muon mass for decay products.}
\end{figure}

\section{Efficiency}
\label{efficiency}

The overall efficiency of the TOF detector can be estimated by means of the track matching efficiency defined as the ratio between the number of tracks reconstructed in the TPC that have produced a detectable signal on TOF and the total number of tracks reconstructed in the TPC. This ratio includes not only the MRPC efficiencies but also dead spaces in the TOF array, particle decays, interaction with materials as well as inefficiencies of the algorithm.\\ 
\indent In Fig. 3 the efficiency of the TOF matching is compared for positive and negative particles in a Monte Carlo (MC) sample and in real data for different $p_t$ values. In the MC the single pad efficiency is 98\% , averaged over the whole pad area and as measured in the test beam. The good match between MC and real data, for this and for other particle samples, indicates that the TOF efficiency is similar to the 98\% used in the MC simulations.

\begin{figure}
\label{fig:eff}
\includegraphics[width=7.5cm, height=7.5cm]{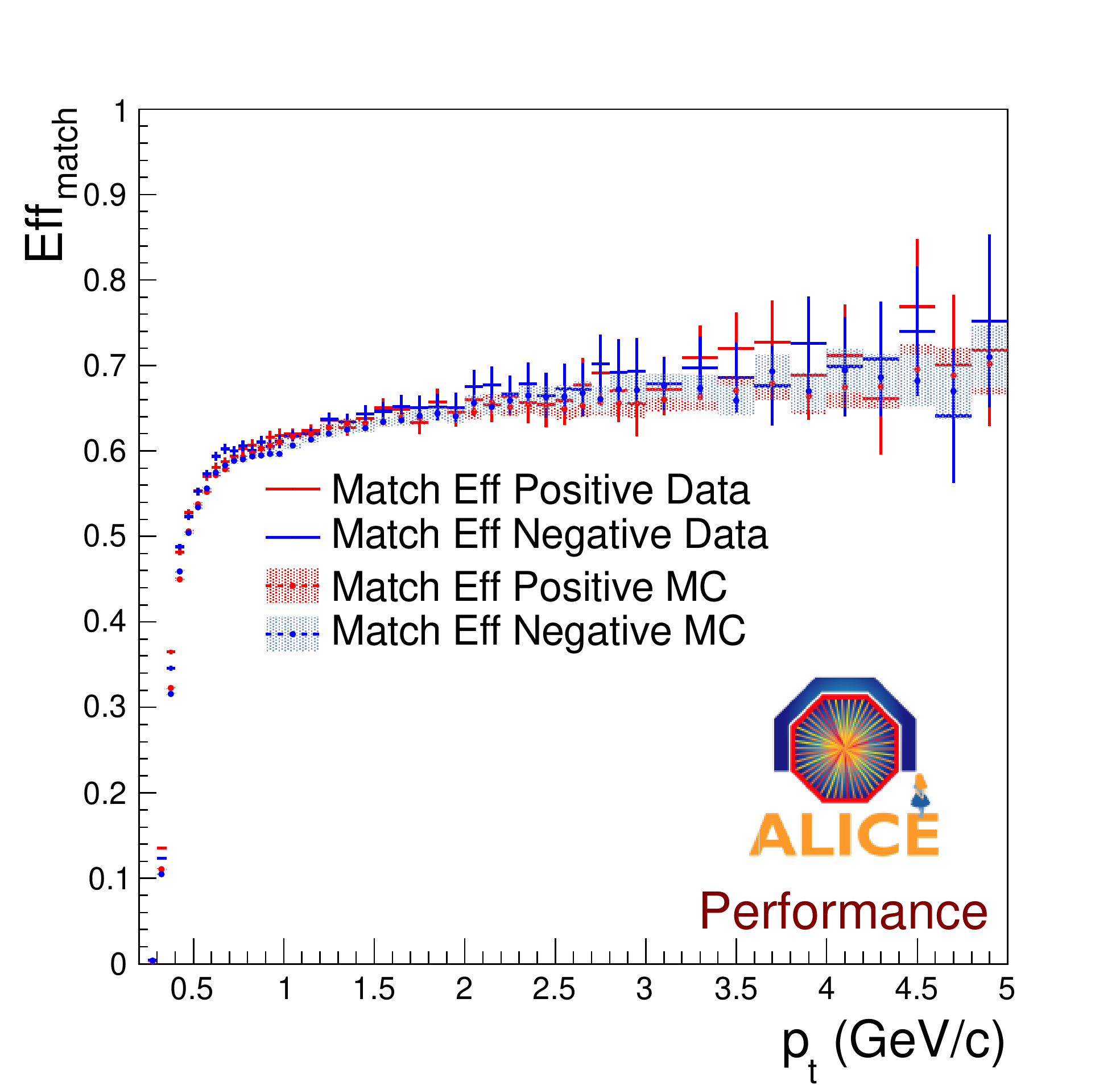}
\caption{TOF track matching efficiency, as a function of $p_t$, compared for positive and negative particles in real data and in MC.}
\end{figure}

\section{TOF PID performance}
\label{pid2}

A time--of--flight measurement is based on the quantity $time_{hit}$ - $t_0$, where $time_{hit}$ is the time measurement made by the TOF detector while $t_0$ is the time of the interaction, measured in ALICE by means of:
\begin{enumerate}[(a)]
\item[A.] a dedicated Cherenkov detector (T0);
\item[B.] TOF detector itself if the amount of tracks reaching the TOF is enough to perform a global time minimization;
\item[C.] the average $t_0$ of LHC fill with resolution $\approx \sigma_Z$/c (finite length of colliding bunches of particles).
\end{enumerate}

The design goal for the ALICE TOF requires a global time resolution of $\approx$ 80 ps. A measurement of this resolution can be performed as in Fig. 3, where the distribution of the time difference between measured and expected time of arrival of pions on TOF is shown, where pion candidates have been selected using TPC information. The plot has been obtained with Pb--Pb collisions at $\sqrt{s_{NN}}$ = 2.76 TeV recorded in November 2010; the result of the fit is the TOF time resolution. The value quoted here includes the intrinsic time resolution of the detector technology plus contributions from electronics (amplification and time--to--digital conversion), uncertainties arising from the distribution of the digital clock from the LHC to the experiments, the spread of the $t_0$ signal (here $\approx$ 10 ps obtained with method B), contributions from the tracking and from calibration. In future the latter can be improved by:

\begin{itemize}
\item[-] single channel time--slewing corrections (finite rise time of the amplifying electronics);
\item[-] time--walk corrections (signal propagation delays on the pick--up pad).
\end{itemize}

\begin{figure}
\label{fig:res}
\includegraphics[width=7.8cm, height=5.6cm]{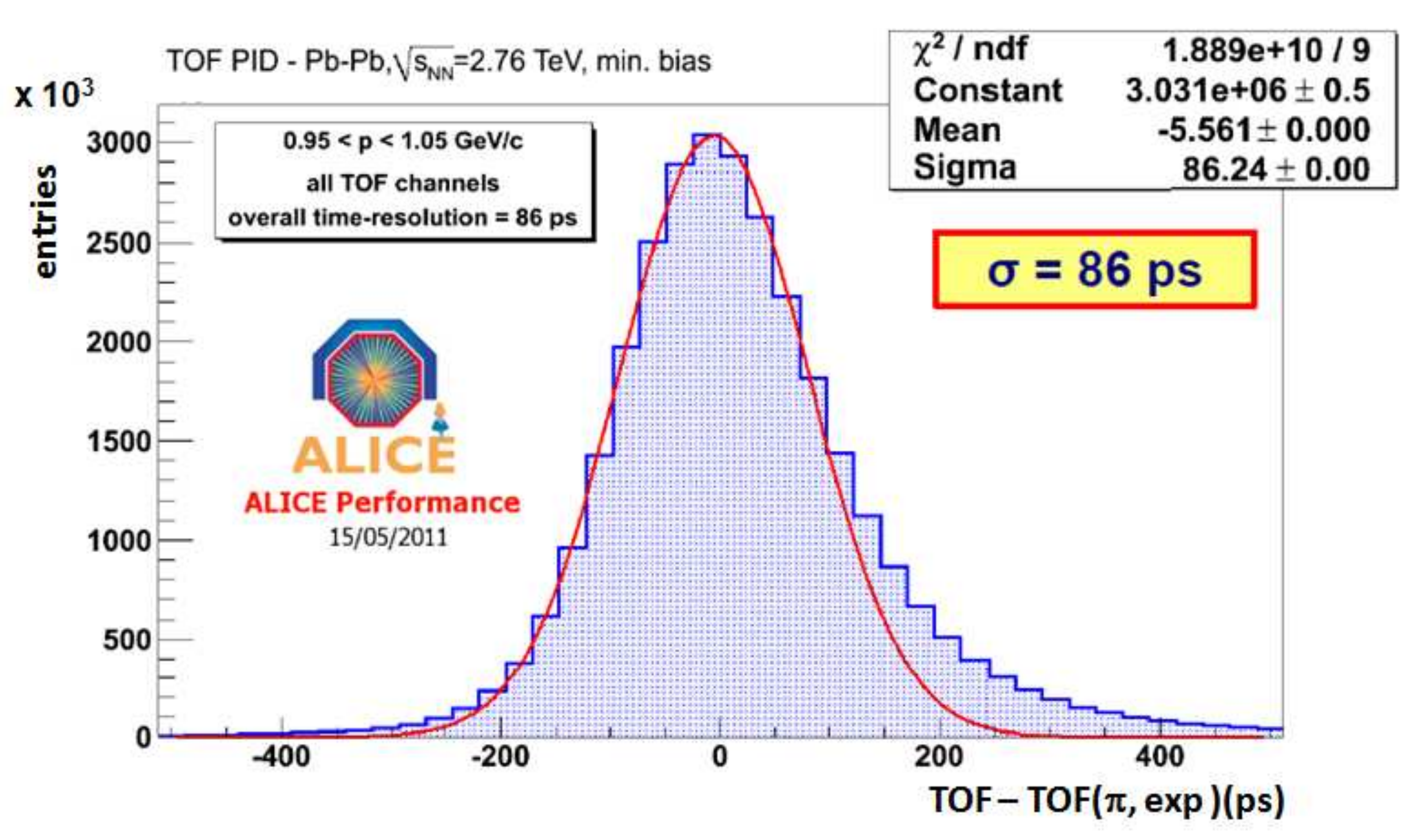}
\caption{TOF PID resolution for selected pions. TOF means the time measured by the TOF detector while TOF($\pi$,exp) is computed during reconstruction by ALICE core central tracking. The red curve shows the gaussian fit to the data.}
\end{figure}

\begin{figure}
\label{fig:sep}
\includegraphics[width=7.7cm, height=5.5cm]{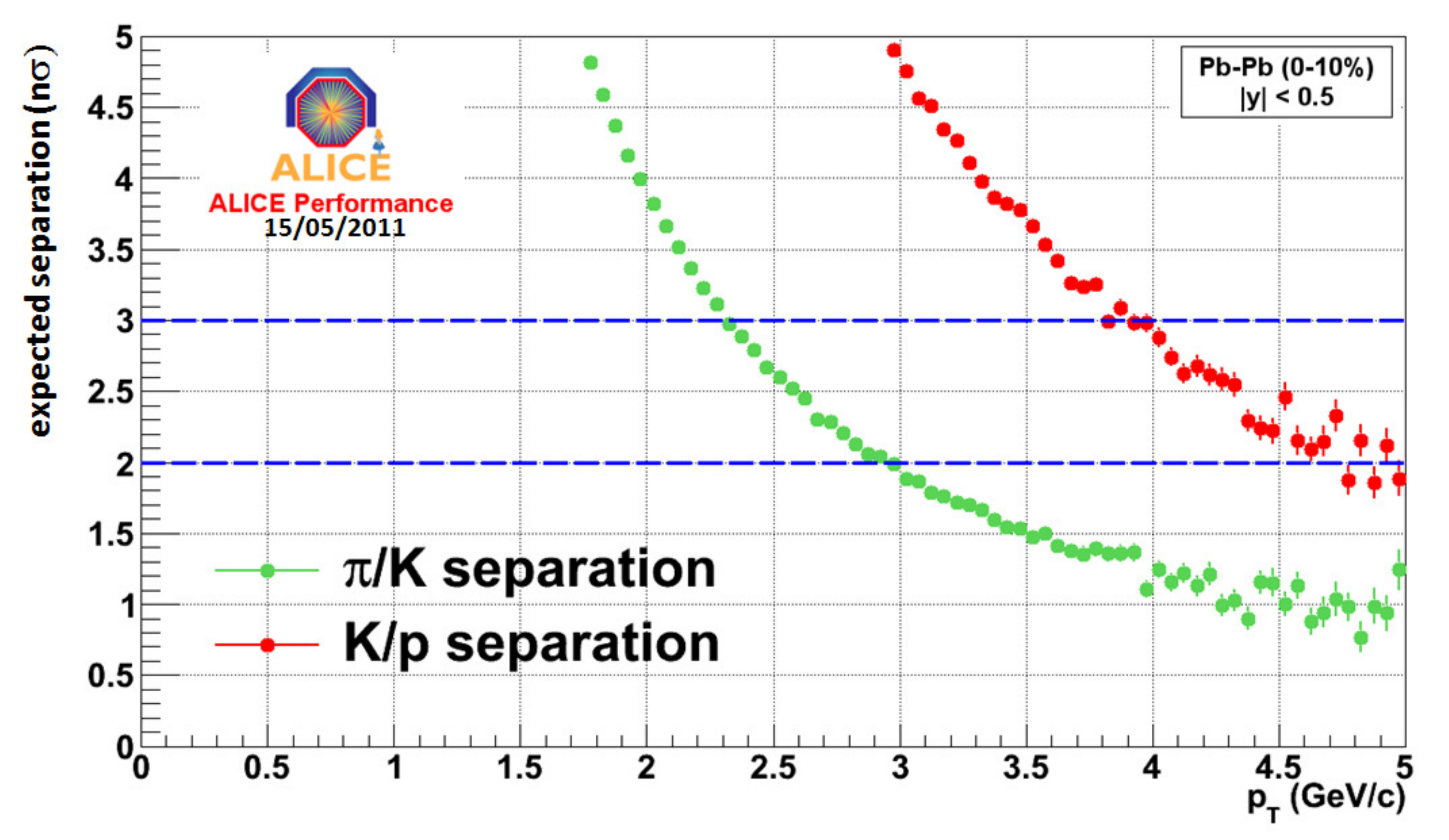}
\caption{Expected separation (nsigma) for $\pi/K$ and $K/p$ as a function of transverse momentum in TOF. Green points $<(t_{K,exp}-t_{\pi,exp})/\sigma(K)>$, red points $<(t_{p,exp}-t_{K,exp})/\sigma(p)>$.}
\end{figure}

In Fig. 4 the expected separation for $\pi/K$ and $K/p$ as a function of transverse momentum in TOF is shown. A two sigma separation is achieved up to 3 GeV for $\pi/K$ and up to 5 GeV for $K/p$ even with a still non--optimal calibration. The plot is based on expected integrated times at TOF smeared according experimental resolution achieved in Pb--Pb central collisions for the different contributions.\\ 
\indent In Fig. 5 is shown the TOF measured $\beta$ as a function of momentum, obtained in Pb--Pb collisions at $\sqrt{s_{NN}}$ = 2.76 TeV; the different particle species are clearly visible.

\begin{figure}
\label{beta-PbPb}
\includegraphics[width=7.8cm, height=5.3cm]{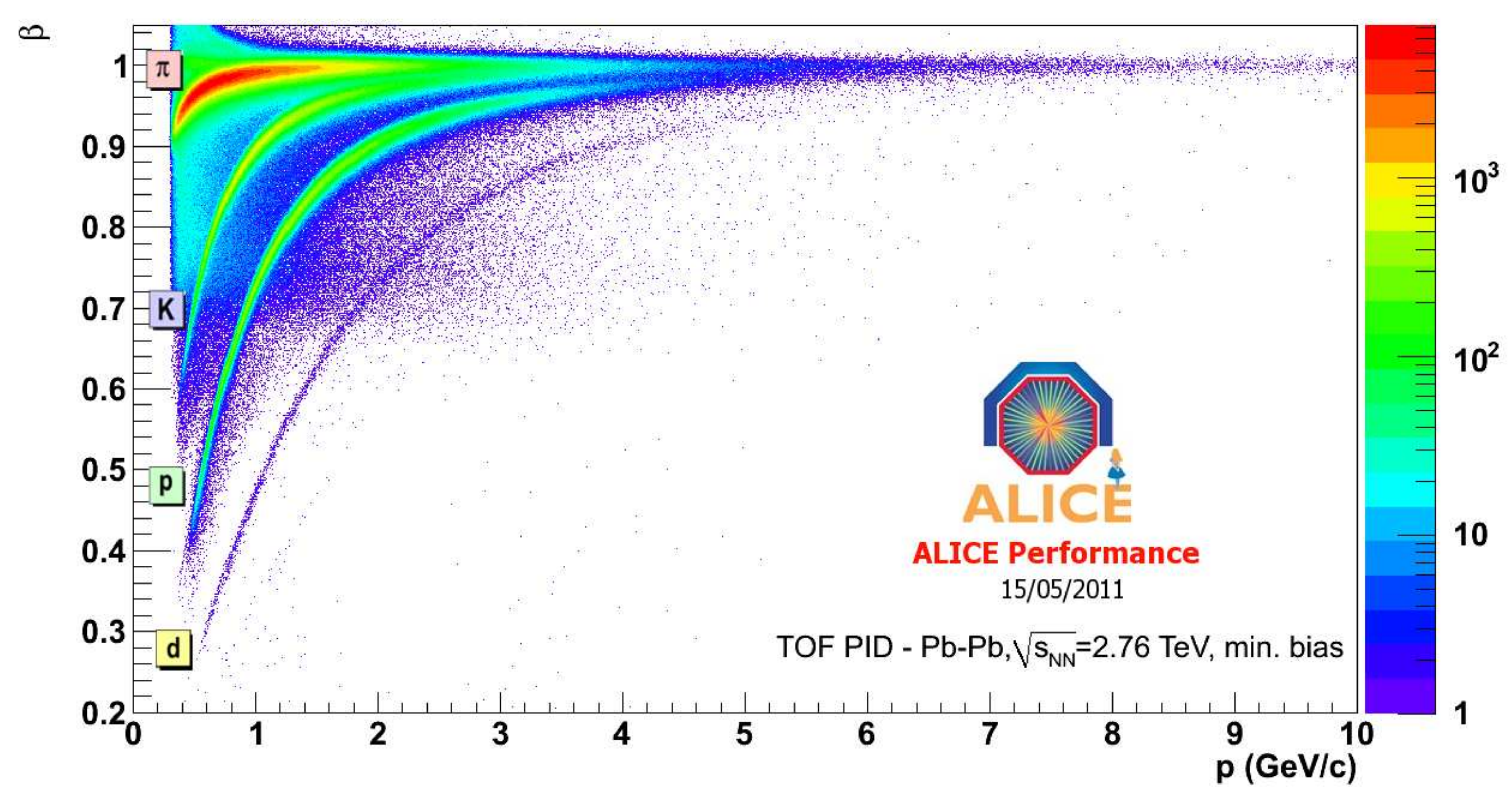}
\caption{TOF measured particle $\beta$ vs. momentum for Pb--Pb collisions at $\sqrt{s_{NN}}$ = 2.76 TeV.}
\end{figure}

\section{Analysis with TOF}
\label{ana}

The TOF information is an important part of the PID analysis which is a prominent feature of the ALICE experiment. Here follow a few examples.\\
\indent Fig. 6 shows the observation of four anti--alpha candidates in ALICE. The light anti--nuclei are sought by looking at the specific energy loss per unit path length in the TPC. Above a rigidity of $p/z$ = 2.3GeV/c however the $dE/dx$ bands from $^3\overline{He}$ and $^4\overline{He}$ are overlapping and the mass calculated with the TOF is needed to separate these two species.\\
\indent The TOF information allows us to extend the $p_t$ reach of the spectra of identified hadrons \cite{Roby}. The TOF also play a fundamental role in the analysis of the anisotropic flow of identified particles \cite{Nofero}.\\
\indent The PID is also used to identify weak decays of strange particles with a sufficiently long lifetime via their characteristic decay topology.\\
\indent Finally, the TOF has been used to improve the separation between electrons and pions in the TPC, by removing the slower hadrons, for a given momentum, with a cut on the particle velocity.

\section{Conclusion}
\label{concl}

The TOF detector has been taking data since the first pp collisions recorded in ALICE in December 2009, with high performance in terms of dark current, noise, efficiency and time resolution. During the 2010 data taking the TOF successfully provided particle identification both for pp and Pb--Pb collisions and many analysis carried out by the ALICE Collaboration are widely using information provided by the TOF detector.

\begin{figure}
\label{fig:antiMatter}
\includegraphics[width=7cm, height=6cm]{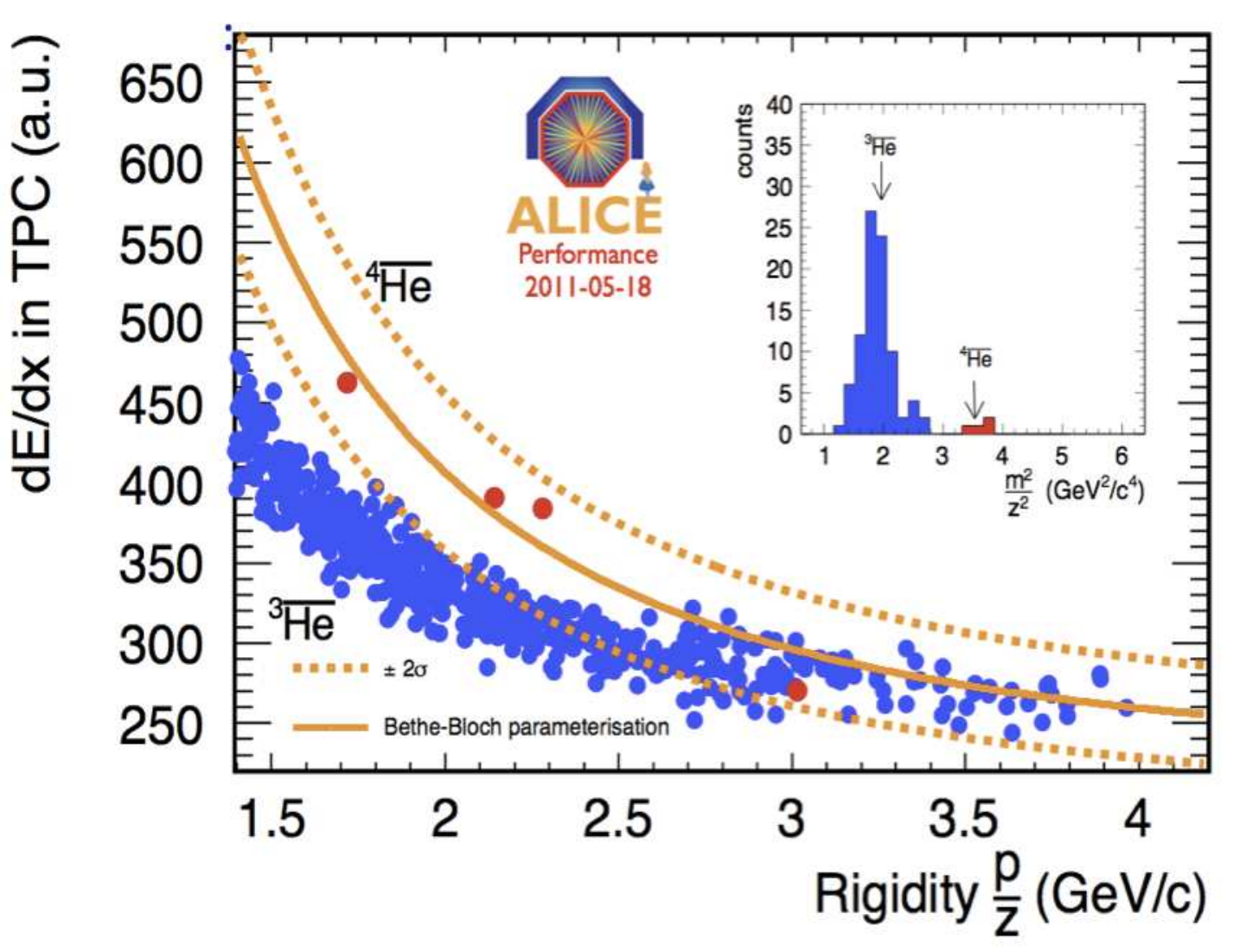}
\caption{Observation of four anti--alpha candidates in ALICE. Above R = 2.3 GeV/c, a clean identification with only TPC is not possible and information from the TOF must be included.}
\end{figure}

\bibliographystyle{model3-num-names}



\end{document}